\documentclass[conference]{IEEEtran}
\IEEEoverridecommandlockouts
\usepackage{url}
\usepackage[acronym]{glossaries}
\usepackage{booktabs}
\usepackage{bbm}
\usepackage{cite}
\usepackage{amsmath,amssymb,amsfonts}
\usepackage{algorithm}
\usepackage{algorithmic}
\usepackage{graphicx}
\usepackage{textcomp}
\usepackage{xcolor}
\usepackage{float}
\usepackage{microtype} 
\usepackage{ragged2e} 
\usepackage{multicol}
\makeglossaries
\newacronym{IAB}{IAB}{Integrated Access and Backhaul}
\newacronym{GNN}{GNN}{Graph Neural Network}
\newacronym{GATv2}{GATv2}{Graph Attention Network v2}
\newacronym{GCN}{GCN}{Graph Convolutional Network}
\newacronym{MDP}{MDP}{Markov Decision Process}
\newacronym{PPO}{PPO}{Proximal Policy Optimization}
\newacronym{DQN}{DQN}{Deep Q-Network}

\newacronym{MIP}{MIP}{Mixed-Integer Programming}
\newacronym{DRL}{DRL}{Deep Reinforcement Learning}
\newacronym{PSO}{PSO}{Particle Swarm Optimization}

\newacronym{VNF}{VNF}{Virtual Network Function}
\newacronym{RAN}{RAN}{Radio Access Network}
\newacronym{O-RUs}{O-RUs}{Open Radio Units}
\newacronym{O-DUs}{O-DUs}{Open Distributed Units}
\newacronym{O-CUs}{O-CUs}{Open Central Units}
\newacronym{RL}{RL}{reinforcement learning}
\newacronym{UAV}{UAV}{Unmanned Aerial Vehicle}
\newacronym{IoT}{IoT}{Internet of Things}
\newacronym{MIMO}{MIMO}{Multiple-Input Multiple-Output}
\newacronym{eMBB}{eMBB}{enhanced Mobile Broadband}
\newacronym{QoS}{QoS}{Quality of Service}
\newacronym{SNR}{SNR}{Signal to Noise Ratio}
\newacronym{CSI}{CSI}{Channel State Information}
\newacronym{DT}{DT}{Digital Twin}
\newacronym{MT}{MT}{Mobile Termination}
\newacronym{DU}{DU}{Distributed Unit}
\newacronym{mmWave}{mmWave}{millimeter-wave}
\newacronym{GIS}{GIS}{Geographic Information System}
\newglossaryentry{LoS}{
  name={LoS},
  description={Line of Sight}
}
\def\BibTeX{{\rm B\kern-.05em{\sc i\kern-.025em b}\kern-.08em
    T\kern-.1667em\lower.7ex\hbox{E}\kern-.125emX}}
\raggedcolumns
\begin{document}

\title{Digital Twin-Assisted Resilient Planning for mmWave IAB Networks via Graph Attention Networks}

\author{
\IEEEauthorblockN{
Jie Zhang\IEEEmembership{, Student Member, IEEE}, 
Mostafa Rahmani Ghourtani\IEEEmembership{, Member, IEEE},
Swarna Bindu Chetty,\\\IEEEmembership{Member, IEEE}, 
Paul Daniel Mitchell\IEEEmembership{, Senior Member, IEEE}, 
and Hamed Ahmadi\IEEEmembership{, Senior Member, IEEE}
}
\IEEEauthorblockA{
\textit{School of Physics, Engineering and Technology}\\
\textit{University of York}\\
York, United Kingdom\\
\
}}

\maketitle
\begin{abstract}
\gls{DT} technology enables real-time monitoring and optimization of complex network infrastructures by creating accurate virtual replicas of physical systems.
In \gls{mmWave} 5G/6G networks, the deployment of \gls{IAB} nodes faces highly dynamic urban environments, necessitating intelligent \gls{DT}-enabled optimization frameworks. Traditional \gls{IAB} deployment optimization approaches struggle with the combinatorial complexity of optimization of joint coverage, connectivity, and resilience, often leading to suboptimal solutions that are vulnerable to the network disruptions. With this consideration, we propose a novel \gls{GATv2}-based reinforcement learning approach for resilient \gls{IAB} deployment in urban \gls{mmWave} networks. Specifically, we formulate the deployment problem as a \gls{MDP} with explicit resilience constraints and employ edge-conditioned \gls{GATv2} to capture complex spatial dependencies between heterogeneous node types and dynamic connectivity patterns. The attention mechanism enables the model to focus on critical deployment locations maximize coverage and ensure fault tolerance through redundant backhaul connections. To address the inherent vulnerability of \gls{mmWave} links, we train the \gls{GATv2} policy using \gls{PPO} with carefully designed balance between coverage, cost, and resilience. Comprehensive simulations across three urban scenarios demonstrate that our method achieves 98.5-98.7\% coverage with 14.3-26.7\% fewer nodes than baseline approaches, while maintaining 87.1\% coverage retention under 30\% link failures—representing 11.3-15.4\% improvement in fault tolerance compared to state-of-the-art methods.
\end{abstract}

\vspace{-0.1in}
\section{Introduction}
\label{sec:introduction}
\gls{DT} technology enables real-time virtual replicas of physical systems, providing unprecedented capabilities for network monitoring, analysis, and optimization \cite{nguyen2021digital}.
In \gls{mmWave} networks, operating in high-frequency bands above 24 GHz, the deployment of \gls{IAB} nodes faces critical challenges due to severe propagation limitations, including high path loss and extreme susceptibility to blockages \cite{rappaport2013millimeter}. Critically, these blockage-prone characteristics make mmWave networks inherently vulnerable to frequent link failures, where temporary obstacles or environmental changes can instantly disconnect users or isolate network segments. In urban environments, these issues are compounded by dense user populations and complex topologies, necessitating a high density of base stations to ensure reliable connectivity. Deploying numerous base stations with traditional wired backhaul, (such as fiber), is prohibitively expensive and logistically challenging. Integrated Access and Backhaul (\gls{IAB}) technology addresses this by enabling nodes to simultaneously provide user access and wireless backhaul to the core network, significantly reducing deployment costs \cite{polese2020integrated}. Optimizing \gls{IAB} node placement is a combinatorial problem that requires balancing coverage, capacity, and resilience against link failures, particularly in dynamic urban settings where blockages and demand fluctuations are common.
\vspace{-0.15in}
\subsection{Related Work and Contributions}
\vspace{-0.1in}
Research on \gls{IAB} deployment optimization primarily focuses on coverage maximization through traditional optimization approaches. Saha et al.~\cite{saha2019integrated} proposed a greedy heuristic algorithm which achieves 92-95\% coverage in small scenarios but requires 30-40\% more infrastructure than optimal solutions due to myopic decision-making that neglects resilience considerations. Polese et al.~\cite{polese2020integrated} formulated \gls{IAB} deployment as \gls{MIP} problem, providing global optimality for networks with fewer than 50 candidate locations but exhibiting exponential complexity $\mathcal{O}(2^N)$ that becomes intractable for realistic urban scenarios exceeding 200 locations. Alzenad et al.~\cite{alzenad2018coverage} employed stochastic geometry for coverage analysis but relied on idealized assumptions including uniform distributions that do not capture urban deployment constraints~\cite{3gpp_tr_38901}.

While coverage optimization has received significant attention, resilience-aware \gls{IAB} deployment remains largely unexplored, despite its critical importance in ensuring network reliability in dynamic environments. Resilience considerations are essential for \gls{IAB} networks due to their multi-hop wireless backhaul nature, where single node failures can cascade through the network topology, potentially disconnecting entire service areas. The wireless nature of \gls{IAB} links makes them inherently susceptible to environmental factors, interference, and physical obstructions that can cause temporary or permanent outages. Moreover, the hierarchical structure of \gls{IAB} networks means that the failure of high-tier nodes has more severe consequences than traditional cellular deployments, making proactive resilience planning crucial rather than reactive recovery approaches.

Existing resilience methods focus primarily on reactive recovery rather than proactive deployment design. Madapatha et al.~\cite{madapatha2020integrated} developed \gls{RL} for dynamic routing in \gls{IAB} networks, achieving a 15\% improvement in failure recovery but assuming a pre-deployed topology. Teymuri et al.~\cite{teymuri2023self} proposed pre-planned optimization for backhaul failures, improving availability by 12\% but suffering from the same scalability limitations as \gls{MIP}-based approaches. Madapatha et al.~\cite{madapatha2021survey} surveyed \gls{IAB} resilience techniques, emphasizing redundant connections but lacking concrete deployment optimization frameworks.

\glspl{GNN} have emerged as promising alternatives for network optimization across various scenarios. Yang et al.~\cite{yang2024graph} developed a generic \gls{GNN}-based node placement achieving 12-18\% improvement over heuristics but lacking resilience constraints. Liu et al.~\cite{liu2022graph} combined \gls{GCN} with deep reinforcement learning for \gls{VNF} orchestration, while Wang et al.~\cite{wang2022joint} applied \glspl{GNN} to \gls{UAV} networks with fundamentally different constraints than terrestrial deployments. Recent advances include specialized architectures for user assignment in mmWave systems~\cite{kunz2024distributed}, multi-head \glspl{GNN} for joint access point selection and beamforming~\cite{zhang2024mhsb}, and heterogeneous \glspl{GNN} for hybrid beamforming optimization~\cite{li2024hgnn,chen2024userbeam}.

Existing \gls{GNN} applications lack \gls{IAB}-specific capabilities, however, particularly resilience-aware deployment optimization. Current approaches focus on coverage without resilience considerations or address resilience reactively rather than through proactive deployment design, failing to model \gls{IAB} networks' unique hierarchical constraints.


\subsection{Research Gap and Contributions}

Most studies focus on coverage or throughput, with few addressing resilience under complex conditions and link failures. \gls{GNN}-based methods show promise but have not been explored for \gls{IAB} deployment. Our contributions are threefold:

1. Resilience-Aware \gls{MDP} Formulation: We formulate \gls{IAB} deployment as a \gls{MDP} with resilience constraints, integrating fault tolerance directly into the optimization problem through penalty-based reward design.

2. Edge-Conditioned \gls{GATv2} Framework: We develop a novel reinforcement learning architecture using \gls{GATv2} with edge-conditioned attention, enabling effective modeling of heterogeneous node types and dynamic link utilization patterns in \gls{IAB} networks.

3. Comprehensive multi-scenario evaluation: We conduct extensive performance analysis across diverse urban topologies with failure testing, demonstrating significant improvements in deployment efficiency (up to 26.7\% node reduction) and fault tolerance (15.4\% better coverage retention) compared to state-of-the-art baselines.

The paper is organized as follows: Section~\ref{sec:sysForm} presents the system model and problem formulation, Section~\ref{sec:simulation} details the simulation setup and results, and Section~\ref{sec:conclusion} concludes the paper.
\vspace{-0.1in}
\section{System Model and Problem Formulation}\label{sec:sysForm}

\subsection{System Model}
\vspace{-0.05in}
We consider an urban mmWave \gls{IAB} network deployment operating in the 60 GHz band, where the service area is discretized into a grid of potential deployment locations, corresponding to existing infrastructure such as lamp posts and utility poles. The network architecture consists of fiber-connected \gls{IAB} donors $\mathcal{I}$, which serve as gateways to the core network, and candidate \gls{IAB} node locations $\mathcal{J}$ that extend coverage via wireless backhaul. Each \gls{IAB} node must maintain at least $m$ independent backhaul connections to ensure network resilience, where $m $ represents the minimum connectivity requirement.

The communication model captures essential mmWave propagation characteristics at 60 GHz. For any link between any two potential nodes (or donor to nodes), the received power is given by:
\begin{equation}
P_r = P_{tx} + G_{\text{total}} - L_{\text{total}}
\label{eq:received_power}
\end{equation}
where $P_{tx}$ denotes transmit power, $G_{\text{total}}$ is the total antenna gain, and $L_{\text{total}}$ represents aggregate propagation loss.

For access links serving user equipment, omni-directional antennas with constant gain $G_0$ are employed:
\begin{equation}
G_{\text{access}} = G_0
\label{eq:access_gain}
\end{equation}

In contrast, backhaul links utilize directional sector antennas to achieve higher gains and interference reduction. The effective backhaul gain is:
\begin{equation}
G_{\text{total}} = \begin{cases}
G_t + G_r, & \text{if } |\theta| \leq \frac{\theta_{\text{HPBW}}}{2} \\
0, & \text{otherwise}
\end{cases}
\label{eq:directional_gain}
\end{equation}
where $G_t$ and $G_r$ are transmit and receive antenna gains, $\theta$ is the angular deviation, and $\theta_{\text{HPBW}}$ is the half-power beamwidth.

The aggregate propagation loss at 60 GHz comprises multiple attenuation components:
\begin{equation}
L_{\text{total}} = L_{path} + L_{atm} + L_{rain}
\label{eq:total_loss}
\end{equation}
accounting for path loss $L_{path}$, atmospheric absorption $L_{atm}$, and rain-induced attenuation $L_{rain}$.

Link feasibility is determined by the \gls{SNR} threshold requirement:
\begin{equation}
\text{SNR} = P_r - N_0 \geq \text{SNR}_{\text{threshold}}
\label{eq:snr_feasibility}
\end{equation}
where thermal noise power $N_0 = -174 + 10\log_{10}(W)$ dBm for bandwidth $W$. 
\vspace{-0.1in}
\subsection{Problem Formulation}
\vspace{-0.1in}
We formulate the resilience-aware \gls{IAB} deployment as a mixed-integer optimization problem that minimizes the number of deployed \gls{IAB} nodes while ensuring coverage, connectivity, and resilience requirements. The problem is non-convex due to binary deployment variables and bilinear capacity constraints.

\subsubsection{Decision Variables}

Let $\mathcal{I}$ denote the set of fiber-connected \gls{IAB} donors and $\mathcal{J}$ the set of candidate \gls{IAB} node locations. The decision variables are:
\begin{itemize}
\item $\alpha_j \in \{0,1\}$: deployment indicator for candidate \gls{IAB} node location $j \in \mathcal{J}$
\item $Y_{pq} \in \{0,1\}$: backhaul link activation from node $p$ to \gls{IAB} node $q$
\item $R_{pq} \geq 0$: traffic flow rate on link $p \rightarrow q$ (Mbps)
\item $U_k \in \{0,1\}$: coverage indicator for grid cell $k \in \mathcal{K}$
\end{itemize}

For all \gls{IAB} donors $i \in \mathcal{I}$, we set $\alpha_i = 1$.

\subsubsection{Coverage Constraints}

The coverage requirement ensures adequate \gls{SNR} levels across the service area:
\begin{equation}
U_k \leq \sum_{i\in\mathcal{I}} C_{ik} + \sum_{j\in\mathcal{J}} C_{jk}\alpha_j, \quad \forall k\in\mathcal{K}
\label{eq:cell_coverage}
\end{equation}
where $C_{ik}, C_{jk} \in \{0,1\}$ indicate whether \gls{IAB} donor $i$ or \gls{IAB} node $j$ provides coverage to cell $k$ with \gls{SNR} $\geq \text{SNR}_{\text{threshold}}$.

The aggregate coverage constraint mandates:
\begin{equation}
\sum_{k\in\mathcal{K}} U_k \geq \theta_{cov} |\mathcal{K}|
\label{eq:total_coverage}
\end{equation}
ensuring at least fraction $\theta_{cov}$ of grid cells receive adequate coverage.

\subsubsection{Resilience and Connectivity Constraints}

Link activation follows logical consistency:
\begin{equation}
Y_{pq} \leq L_{pq}\alpha_p\alpha_q, \quad \forall p \in \mathcal{I}\cup\mathcal{J}, q \in \mathcal{J}, p \neq q
\label{eq:link_activation}
\end{equation}
where $L_{pq} \in \{0,1\}$ indicates link feasibility based on \gls{SNR} requirements from the communication model.

The resilience constraint ensures redundant connectivity:
\begin{equation}
\sum_{p \in \mathcal{I}\cup\mathcal{J}, p \neq j} Y_{pj} \geq m \cdot \alpha_j, \quad \forall j\in\mathcal{J}
\label{eq:resilience_constraint}
\end{equation}
requiring each deployed \gls{IAB} node ($\alpha_j = 1$) to maintain at least $m$ active backhaul connections for fault tolerance.

\subsubsection{Capacity and Flow Constraints}

Link capacity constraints reserve bandwidth for failure recovery:
\begin{equation}
R_{pq} \leq (1-\beta) C_{pq} Y_{pq}, \quad \forall p \in \mathcal{I}\cup\mathcal{J}, q \in \mathcal{J}, p \neq q
\label{eq:link_capacity}
\end{equation}
where $C_{pq}$ is the physical link capacity (Mbps) and $\beta \in [0,1]$ reserves fraction $\beta$ of link capacity for traffic rerouting 
during link failures. This ensures that when primary paths fail, backup routes have 
sufficient capacity.

\gls{IAB} donor capacity constraints limit fiber backhaul usage:
\begin{equation}
\sum_{j\in\mathcal{J}} R_{ij} Y_{ij} + R_o A_i \leq F_i, \quad \forall i\in\mathcal{I}
\label{eq:donor_capacity}
\end{equation}
where $F_i$ is \gls{IAB} donor $i$'s fiber capacity, $A_i > 0$ represents local access demand, and $R_o > 1$ accounts for protocol overhead including MAC layer framing, 
PHY layer pilot symbols, routing protocol signaling, and retransmission mechanisms.

Flow conservation at \gls{IAB} nodes ensures traffic balance:
\begin{equation}
\sum_{p \in \mathcal{I}\cup\mathcal{J}, p \neq j} R_{pj} Y_{pj} \geq R_o \left( A_j \alpha_j + \sum_{n\in\mathcal{J}, n \neq j} R_{jn} Y_{jn} \right), \quad \forall j \in \mathcal{J}
\label{eq:flow_conservation}
\end{equation}
where inbound traffic (left side) must satisfy local access demand $A_j$ and outbound forwarding requirements (right side), both scaled by overhead factor $R_o$.

\subsubsection{Optimization Objective}

The objective minimizes the number of deployed \gls{IAB} nodes:
\begin{equation}
\min \sum_{j\in\mathcal{J}} \alpha_j
\label{eq:objective}
\end{equation}
subject to constraints \eqref{eq:cell_coverage}-\eqref{eq:flow_conservation}.

This formulation ensures optimal \gls{IAB} node placement while maintaining network resilience through redundant connectivity and capacity reservation for failure recovery.
\subsection{Graph Representation}
\label{sec:graphForm}
To enable \gls{GNN}-based optimization, we transform the mixed-integer formulation into a heterogeneous attributed digraph $\mathcal{G} = (\mathcal{V}, \mathcal{E}, \mathbf{X}, \mathbf{E}, \mathbf{g})$ that captures network topology and deployment state.

\paragraph{Vertices} $\mathcal{V} = \mathcal{I} \cup \mathcal{J}$ comprise \gls{IAB} donors ($i \in \mathcal{I}$) and candidate \gls{IAB} node locations ($j \in \mathcal{J}$).
\paragraph{Directed edges} $\mathcal{E} = \{(p, q) \mid L_{pq} = 1, p \in \mathcal{I} \cup \mathcal{J}, q \in \mathcal{J}, p \neq q\}$ represent feasible mmWave backhaul links satisfying \gls{SNR} requirements.

\paragraph{Node Features $\mathbf{X} \in \mathbb{R}^{|\mathcal{V}| \times d_v}$}
Each vertex $v \in \mathcal{V}$ has feature vector:
\[
  \mathbf{x}_v = \left[ \alpha_v, \frac{A_v}{A_{\max}}, \frac{N_v}{m}, \mathbbm{1}_{\{v \in \mathcal{I}\}} \right]
\]
where $\alpha_v \in \{0,1\}$ is the deployment status, $\frac{A_v}{A_{\max}}$ is normalized access demand (with $A_{\max} = \max_{v \in \mathcal{V}} A_v$ from traffic analysis), $\frac{N_v}{m}$ represents resilience ratio (current connections over requirement $m$), and $\mathbbm{1}_{\{v \in \mathcal{I}\}}$ indicates \gls{IAB} donor type. Normalization prevents gradient instability caused by different feature scales.

\paragraph{Edge Features $\mathbf{E} \in \mathbb{R}^{|\mathcal{E}| \times d_e}$}
Each directed link $(p, q) \in \mathcal{E}$ carries:
\[
  \mathbf{e}_{pq} = \left[ \frac{C_{pq}}{C_{\max}}, \frac{R_{pq}}{C_{pq}}, L_{pq} \right]
\]
where $\frac{C_{pq}}{C_{\max}}$ is normalized link capacity (with $C_{\max} = \max_{(p,q) \in \mathcal{E}} C_{pq}$ from strongest feasible link), $\frac{R_{pq}}{C_{pq}}$ represents current utilization ratio, and $L_{pq} = 1$ confirms link feasibility.

\paragraph{Global Features $\mathbf{g} \in \mathbb{R}^{4}$}
Graph-level parameters:
\[
  \mathbf{g} = \left[ \theta_{cov}, m, \beta, R_o \right]
\]
encoding coverage target, resilience requirement, capacity headroom, and protocol overhead.

This representation enables the \gls{GNN} to learn spatial dependencies while reasoning about deployment constraints, resilience requirements, and capacity utilization during the optimization process.

\subsection{Markov Decision Process Formulation}
\vspace{-0.03in}
We formulate the \gls{IAB} deployment problem as a \gls{MDP} $(\mathcal{S}, \mathcal{A}, \mathcal{P}, \mathcal{R}, \gamma)$ to enable reinforcement learning optimization:

\paragraph{State Space $\mathcal{S}$}
The state $s_t \in \mathcal{S}$ at step $t$ is represented by the attributed graph $\mathcal{G}_t = (\mathcal{V}, \mathcal{E}, \mathbf{X}_t, \mathbf{E}_t, \mathbf{g})$, where node and edge features encode current deployment status, traffic flows, and connectivity patterns.

\paragraph{Action Space $\mathcal{A}$}
At each step, the agent selects action $a_t \in \mathcal{A} = \{0\} \cup \{j \mid j \in \mathcal{J}, \alpha_j = 0\}$ to either deploy an \gls{IAB} node at candidate location $j$ ($a_t = j$) or maintain current configuration ($a_t = 0$).

\paragraph{Reward Function $\mathcal{R}$}
The reward function balances coverage improvement against deployment costs and resilience violations:
\[
r_t = \kappa_1 \Delta U_{cov} - \kappa_2 \alpha_{deploy} - \kappa_3 N_{vulnerable}
\]
where $\Delta U_{cov}$ represents coverage percentage improvement, $\alpha_{deploy} \in \{0,1\}$ indicates new \gls{IAB} node deployment, and $N_{vulnerable}$ penalizes nodes violating the resilience constraint $m$. The coefficients $\kappa_1, \kappa_2, \kappa_3$ control the relative importance of coverage gains versus deployment costs and constraint violations.

\begin{algorithm}[htbp]
\caption{GATv2-PPO for Resilient IAB Deployment}
\label{GATv2-PPO}
\begin{algorithmic}[1]
\REQUIRE Donor configuration $\mathcal{D}$, candidate locations $\mathcal{J}$, training episodes $E$
\ENSURE Optimized deployment policy $\pi^*_\theta$

\FOR{$e = 1$ to $E$}
    \STATE Initialize environment with fixed donor topology $\mathcal{D}$
    \STATE Reset deployment state $\mathbf{s}_0$ and coverage metrics
    
    \WHILE{coverage threshold $\theta_{cov}$ not achieved}
        \STATE Construct heterogeneous graph $\mathcal{G}_t = (\mathcal{V}, \mathcal{E}, \mathbf{X}_t, \mathbf{E}_t)$
        \STATE Apply GATv2 encoding: $\mathbf{H}^{(L)} = \text{GATv2}(\mathbf{X}_t, \mathbf{E}_t, \mathcal{E})$
        \STATE Sample deployment action: $a_t \sim \boldsymbol{\pi}_t$
        \STATE Execute action $a_t$ and observe transition $(s_t, a_t, r_t, s_{t+1})$
        \STATE Compute reward $r_t$ with resilience penalties
        \STATE Store experience $(\mathcal{G}_t, a_t, r_t, \mathcal{G}_{t+1}, d_t)$ in replay buffer
        \STATE Update graph state $\mathcal{G}_{t+1}$ based on deployment changes
    \ENDWHILE
\ENDFOR

\RETURN Trained policy $\pi^*_\theta$ achieving resilient network deployment
\end{algorithmic}
\end{algorithm}

\paragraph{GATv2-based Policy Architecture}
The policy $\pi_\theta(a|s)$ employs a GATv2 encoder with edge-conditioned attention to process the graph state:
\[
\mathbf{h}_v^{(l+1)} = \text{GATv2}^{(l)}(\mathbf{h}_v^{(l)}, \{\mathbf{h}_u^{(l)}, \mathbf{e}_{uv}\}_{u \in \mathcal{N}(v)})
\]
where $\mathbf{h}_v^{(l)}$ represents node embeddings at layer $l$, and edge features $\mathbf{e}_{uv}$ condition the attention mechanism. A pointer-based actor network computes deployment probabilities over valid candidate locations, while a critic network estimates state values for \gls{PPO} training.
\vspace{-0.1in}
\begin{table}[!t]
\centering
\caption{Key Symbols and Definitions}
\label{tab:notation}
\begin{tabular}{ll}
\toprule
\textbf{Symbol} & \textbf{Definition} \\ \midrule
$\mathcal{I}, \mathcal{J}, \mathcal{K}$ & \gls{IAB} donor set, candidate \gls{IAB} node set, grid cell set \\
$\alpha_j$ & Binary deployment variable for \gls{IAB} node $j$ \\
$Y_{pq}$ & Binary link activation variable \\
$U_k$ & Binary coverage indicator for cell $k$ \\
$C_{ik}, C_{jk}$ & Coverage indicators (0/1) \\
$L_{pq}$ & Link feasibility indicator (0/1) \\
$R_{pq}$ & Traffic flow on link $p \rightarrow q$ [Mbps] \\
$C_{pq}$ & Physical capacity of link $p \rightarrow q$ [Mbps] \\
$A_i, A_j$ & Access demand [Mbps] \\
$F_i$ & Fiber capacity of \gls{IAB} donor $i$ [Mbps] \\
$m$ & Minimum resilience degree (inbound links) \\
$R_o$ & Protocol overhead factor ($> 1$) \\
$\beta$ & Backup capacity fraction for resilience \\
$\theta_{cov}$ & Target coverage fraction \\
$N_v$ & Number of active inbound connections to node $v$ \\
\bottomrule
\end{tabular}
\end{table}
\section{Simulation and Results}
\label{sec:simulation}
\vspace{-0.1in}
\subsection{Simulation setup}
\vspace{-0.05in}
The simulation environment models urban mmWave \gls{IAB} deployment across 1×1 km$^2$ service areas with 400 potential node locations distributed at 50m intervals to mimic realistic urban infrastructure density, corresponding to lamp posts and utility poles. This grid-based approach ensures fair algorithm comparison while maintaining practical relevance. Three deployment scenarios are used to evaluate algorithm performace: Pentagon (5 donors in pentagonal setting), Five-Dice (donors positioned as dice-5 pattern), and Vertical (linear arrangement). Table~\ref{tab:simulation_params} summarizes the key simulation parameters used throughout the evaluation.
\begin{table}[!t]
\centering
\caption{Simulation Parameters and Model Configuration}
\label{tab:simulation_params}
\begin{tabular}{llc}
\toprule
\textbf{Parameter} & \textbf{Value} & \textbf{Reference} \\
\midrule
Transmit power $P_{tx}$ & 30 dBm & \cite{itu_m2410} \\
Antenna gain $G_t, G_r$ & 25 dBi & \cite{itu_m2410} \\
Noise figure & 7 dB & \cite{itu_m2410} \\
Operating frequency & 60 GHz & \cite{itu_m2410} \\
\gls{SNR} threshold & 10 dB & \cite{3gpp_tr_38901} \\
\midrule
Coverage threshold $\theta_{cov}$ & 98\% & -- \\
Backup capacity fraction for resilience $\beta$ & 0.2 & -- \\
Overhead $R_o$ & 1.2 & -- \\
Resilience parameter $m$ & 2 & -- \\
Reward coefficients $\kappa_1, \kappa_2, \kappa_3$ & 4.0, 0.2, 0.5 & -- \\
\midrule
Graph encoder & 2-layer GATv2 & -- \\
Hidden units & 64 per layer & -- \\
Attention heads & 8 & -- \\
Learning rate & $3 \times 10^{-4}$ & -- \\
Batch size & 32 & -- \\
Discount factor $\gamma$ & 0.99 & -- \\
Clip ratio & 0.2 & -- \\
Training episodes & 8000 & -- \\
\bottomrule
\end{tabular}
\end{table}

To address the resilience-aware \gls{IAB} deployment challenge, we propose \gls{GATv2} with edge-conditioned attention mechanism detailed in Algorithm \ref{GATv2-PPO} to capture spatial dependencies and connectivity constraints inherent in \gls{IAB} networks. The model's ability to process heterogeneous node types (\gls{IAB} donors vs. candidate nodes) and dynamic edge features (link capacity, utilization) makes it particularly suited for modeling complex relationships between deployment decisions, network topology, and resilience requirements. The model is trained using Proximal Policy Optimization (\gls{PPO}) within the \gls{MDP} framework, enabling the agent to balance coverage objectives against deployment costs and constraint violations.
Model performance is benchmarked against two established approaches:
\begin{itemize}
\item \textbf{Greedy Heuristic}: A greedy algorithm \cite{info15010019} that sequentially deploys nodes by maximizing immediate coverage improvement under data rate constraints and solved by Gurobi solver. Although fast and interpretable, its myopic nature leads to suboptimal long-term performance.
\item \textbf{Dueling DQN with GCN}: Deep reinforcement learning approach using a Graph Convolutional Network backbone with experience replay and target networks~\cite{10154469}.
\end{itemize}

\begin{figure}[!t]
\centering
\includegraphics[width=\columnwidth]{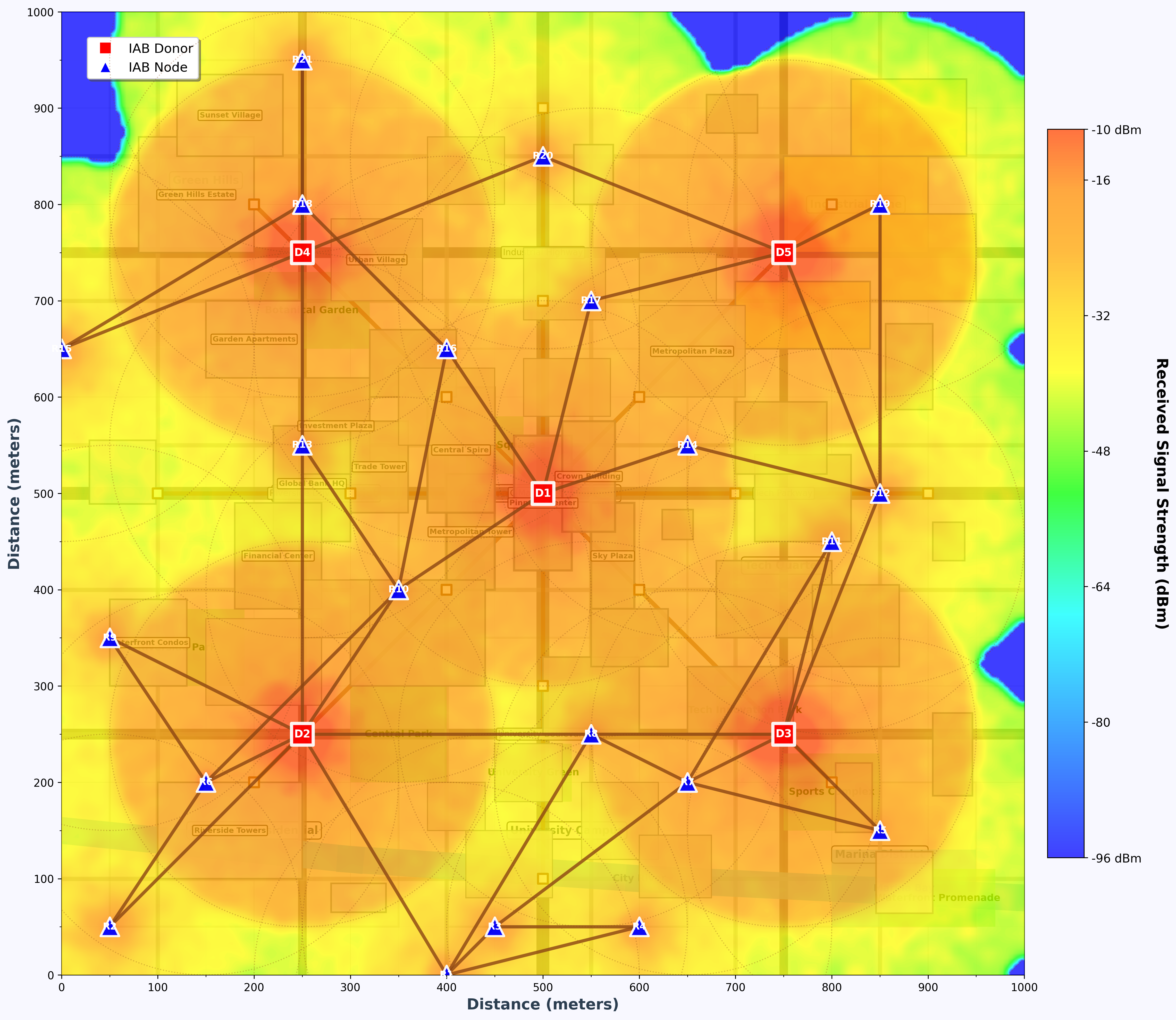}
\caption{\gls{GATv2}-optimized IAB network deployment showing Five-Dice donor configuration (red squares D1-D5), deployed nodes (blue triangles), and signal strength heat map. Black lines indicate backhaul connections forming a resilient mesh topology with redundant paths.}
\label{fig:network_deployment}
\end{figure}

\begin{figure}[!t]
\centering
\includegraphics[width=\columnwidth]{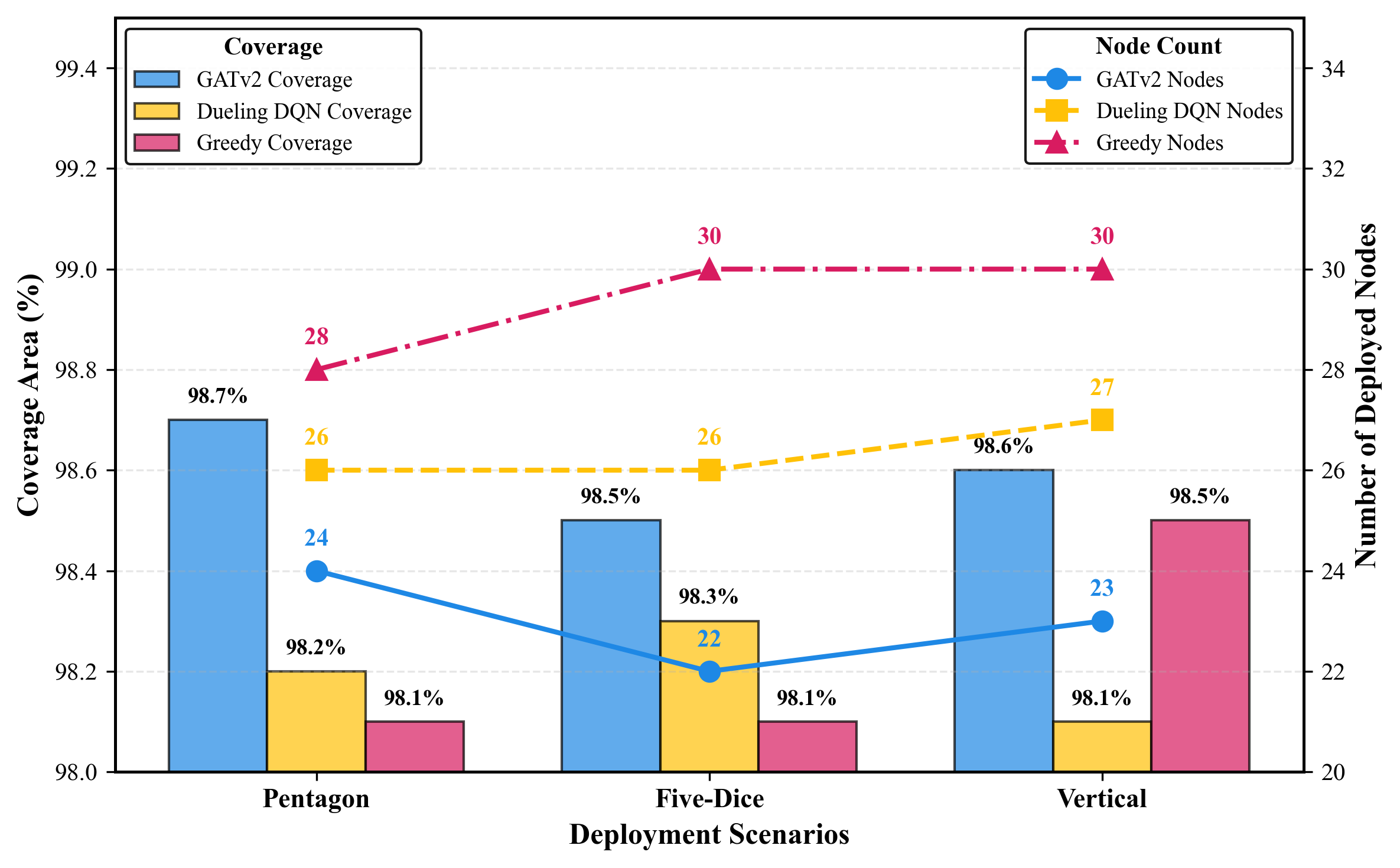}

\caption{Multi-scenario deployment efficiency comparison across Pentagon, Five-Dice, and Vertical donor configurations.}
\label{fig:multi_scenario}

\end{figure}

\subsection{Result analysis}
\vspace{-0.05in}
Fig.~\ref{fig:network_deployment} demonstrates the spatial deployment achieved by \gls{GATv2} in the Five-Dice configuration. The heat map shows coverage zones from -96 dBm to -10 dBm, with deployed nodes (blue triangles) creating overlapping high-\gls{SNR} areas that eliminate coverage gaps between donors D1-D5. The mesh connectivity pattern (black lines) ensures each node maintains multiple backhaul connections, satisfying the $m=2$ redundancy constraint through \gls{GATv2}'s learned policy that balances coverage and resilience requirements.
Building upon this spatial optimization capability, Fig.~\ref{fig:multi_scenario} quantifies deployment efficiency across three donor configurations. \gls{GATv2} consistently outperforms baselines: Pentagon scenario achieves 98.7\% coverage with 24 nodes versus 26 (DQN, 98.2\%) and 28 (Greedy, 98.1\%); Five-Dice requires only 22 nodes for 98.5\% coverage, representing 26.7\% reduction compared to Greedy's 30 nodes; Vertical uses 23 nodes versus 27 (DQN) and 30 (Greedy) for 98.6\% coverage. These improvements result from \gls{GATv2}'s attention mechanism capturing multi-hop spatial dependencies for global topology optimization, surpassing greedy methods' myopic decisions and standard GCNs' limited graph representation.

\begin{figure}[!t]
\centering
\includegraphics[width=\columnwidth]{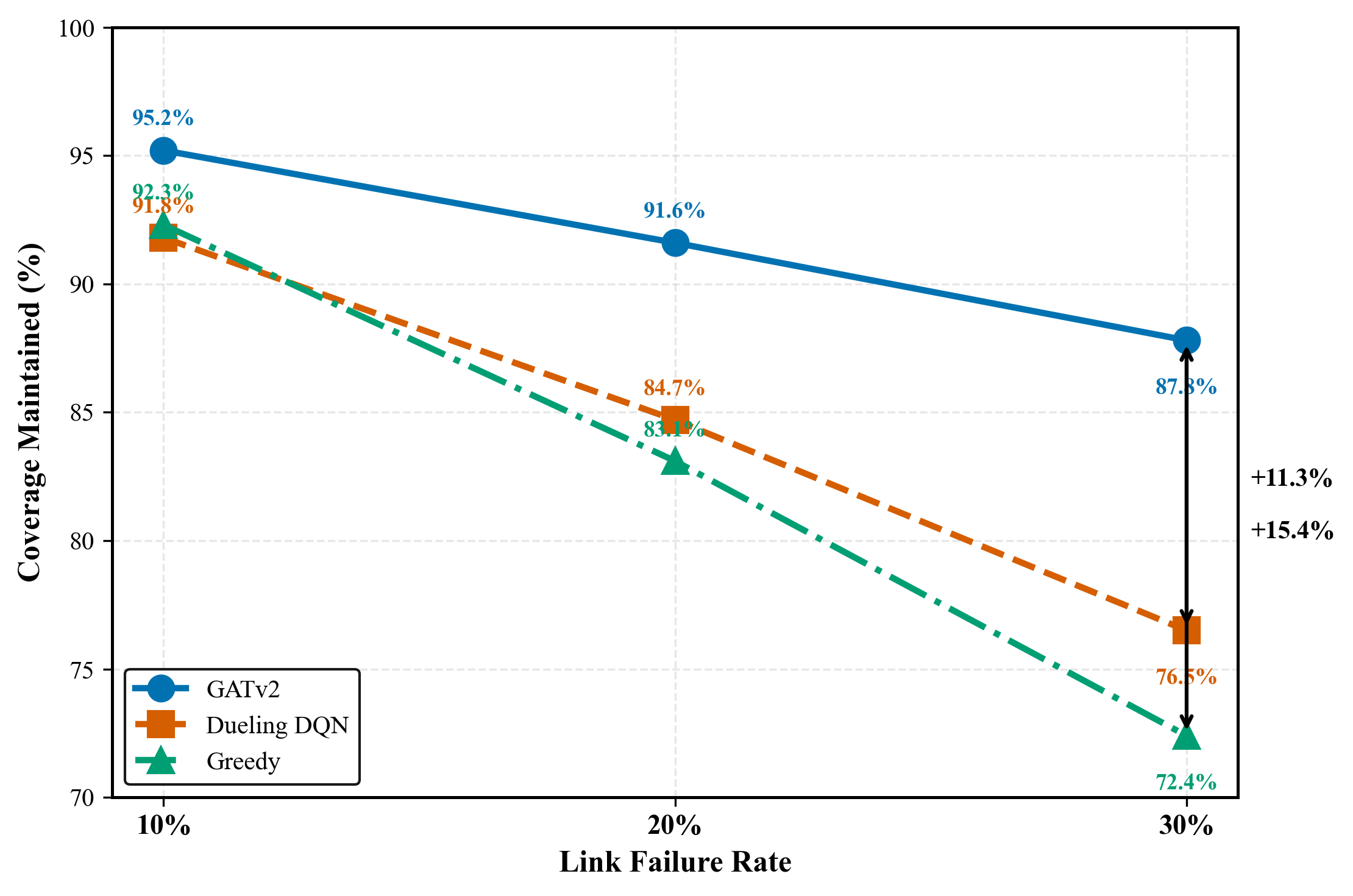}
\caption{Network resilience under progressive link failures, and progressive failure tests randomly disable 10\%, 20\%, and 30\% of backhaul links to simulate mmWave blockage events.}
\label{fig:resilience_analysis}
\end{figure}
Beyond deployment efficiency, the resilience-aware design proves critical under network stress conditions. To evaluate fault tolerance, we conduct link failure simulation using random failure models. The methodology randomly disables 10\%, 20\%, and 30\% of backhaul links, with each failure rate tested across 100 independent trials. We measure coverage retention as the network robustness indicator.
Fig.~\ref{fig:resilience_analysis} demonstrates that \gls{GATv2} maintains superior performance across all failure scenarios: 95.2\% coverage retention at 10\% failure versus 91.3\% (DQN) and 92.3\% (Greedy); 91.6\% at 20\% failure versus 84.7\% (DQN) and 83.1\% (Greedy); 87.1\% at 30\% failure versus 76.6\% (DQN) and 72.4\% (Greedy). The 11.3-15.4\% performance advantage stems from the explicit $m$ connectivity constraint in the \gls{MDP} formulation, preventing single-connected deployments that cause cascading failures in baseline methods.
\vspace{-0.15in}
\subsection{Complexity Analysis}
\vspace{-0.05in}
The computational complexity of GATv2-based deployment is analyzed for a network with $N$ candidate nodes and $E$ potential backhaul links. The graph attention mechanism requires $\mathcal{O}(E \cdot d_h)$ operations per attention head, where $d_h$ is the hidden dimension. With 2 attention heads and 2 GATv2 layers, the encoding complexity is $\mathcal{O}(4E \cdot d_h)$. The actor network performs $\mathcal{O}(|\mathcal{A}| \cdot d_h)$ operations for action selection over valid action set $\mathcal{A}$.

For our urban scenarios with $N=400$ nodes and average node degree 8, $E \approx 1600$. With $d_h=32$, total complexity per episode is $\mathcal{O}(51200 + 32|\mathcal{A}|)$. This linear scaling contrasts favorably with mixed-integer programming approaches exhibiting exponential complexity $\mathcal{O}(2^N)$ for binary placement variables.

The PPO training complexity is $\mathcal{O}(B \cdot T \cdot C_{forward})$ per update, where $B=32$ is batch size, $T$ is trajectory length, and $C_{forward}$ is the forward pass cost. Memory requirements scale as $\mathcal{O}(N \cdot d_h + E \cdot d_e)$ for node and edge embeddings, remaining manageable for large-scale deployments.

\section{Conclusion}
\label{sec:conclusion}
\vspace{-0.1in}
This paper introduced a novel \gls{GATv2}-based \gls{RL} approach for resilient \gls{mmWave} \gls{IAB} network deployment. Our resilience-aware \gls{MDP} formulation enables optimal deployment patterns balancing coverage, efficiency, and fault tolerance. Experimental results demonstrate superior performance: over 98\% coverage with 14.3-26.7\% fewer nodes than baselines, 87.1\% coverage retention under 30\% link failures (11.3-15.4\% improvement over competing methods), and linear computational scaling suitable for large deployments. The edge-conditioned \gls{GATv2} architecture effectively captures \gls{mmWave} spatial dependencies while preventing vulnerable single-connected deployments. These results validate \gls{GNN}-based approaches for complex network optimization and suggest promising extensions to O-RAN architectures and 6G orchestration, particularly through integration with \gls{DT} technology for real-time adaptive deployment in next-generation wireless systems.
\vspace{-0.15in}

\section*{Acknowledgment}
\vspace{-0.1in}
This work was supported in part by the Engineering and Physical Sciences Research Council United Kingdom (EPSRC), Impact Acceleration Accounts (IAA) (Green Secure and Privacy Aware Wireless Networks for Sustainable Future Connected and Autonomous Systems) under Grant EP/X525856/1, EPSRC CHEDDAR: Communications Hub for Empowering Distributed ClouD Computing Applications and Research under grant numbers EP/Y037421/1 and EP/X040518/1, and Department of Science, Innovation and Technology, United Kingdom, under Grants Yorkshire Open-RAN (YORAN) TS/X013758/1 and RIC Enabled (CF-)mMIMO for HDD (REACH) TS/Y008952/1.

\bibliographystyle{IEEEtran}

\end{document}